\documentclass[10pt]{article}
\usepackage{amsthm}
\title{Zone Theorem for Arrangements in three dimensions}

\author{Sanjeev Saxena\thanks{E-mail: ssax@iitk.ac.in}\\
Dept. of Computer Science and
Engineering,\\ Indian Institute of Technology,\\
Kanpur, INDIA-208 016}

\date{\today}
\newtheorem{theorem}{Theorem}
\begin{document}
\maketitle

\subsection*{\centering{Abstract}}

In this note, a simple description of zone theorem in three dimensions
is given. Arrangements in three dimensions are useful for constructing
higher-order Voronoi diagrams in plane. An elementary and very
intuitive treatment of this result is also given.

{{Keywords:}} Computational Geometry, Zone Theorem, Arrangements,
$k$-Nearest Neighbours

\section*{1. Introduction}

Zone theorem is important in analysing incremental algorithms for
constructing arrangements.  Most popular text books of Computational
Geometry (see e.g. \cite{R1,R2}) describe zone theorem for
Arrangements in two dimensions. Specialised books like \cite{R3,R4}
describe zone theorem for hyperplanes in $d$-dimensions. Three
dimensional arrangements are useful for constructing higher order
Voronoi diagrams in plane \cite{R1, R5, R4, R3,RB}. 
An elementary and very
intuitive treatment of this is given in Section~4.
Proofs of Zone
theorem in higher dimensions use Euler's relation: $\sum_{i=0}^d
(-1)^i F_i\geq 0$ \cite{R3,R6,RA}.  As most students of Computational
Geometry are not familiar with these result, only zone theorem in two
dimensions is taught in most Computational Geometry courses. In this
note, a proof of zone theorem in three dimensions which can be easily
taught in Computational Geometry courses is described. The proof uses
zone theorem in two dimensions \cite{R1, R2,R3,RA,ipl}. The proof is a
essentially a simplified version of proof given by Edelsbrunner,
Seidel and Sharir\cite{RA}.

\section*{2. Definitions and basic properties}

Arrangement in two dimensions is basically a set of $n$ lines
(infinite lines and not segments), and in three dimensions of $n$
planes. We will assume that lines in two dimensions and planes in
three dimensions are in general position. Thus, in two dimension no
two lines are parallel and no three lines meet in a single point
\cite{R1,R2,RA}. Similarly, in three dimensions we will assume
that
\begin{itemize}
\item No two planes are parallel. Thus, each pair of planes meet in a
line. And any three planes in a point.
\item No three planes intersect in  a common line and no four planes
in a (common) point.
\end{itemize}
Set of lines, in two dimensions, will partition the plane into regions
called faces. And set of planes in three dimensions will partition the
space into regions, which we will call cells.

Let $C$ be any bounded cell.  As planes are in general position, each
point or vertex $v$ in arrangement is determined by three planes.
Thus, there will be three edges of $C$ incident at any vertex $v$ of
$C$. Moreover, each edge of $C$ is determined by two vertices of $C$.
If $|V_C|$ is the number of vertices of $C$ and $|E_C|$ is the number
of edges of $C$, then $2|E_C|=3|V_C|$. 

As $C$ is on one side of each plane, $C$ will be a convex polytope
($3$-dimensional analogue of polygon). The set of edges, vertices and
faces on boundary of $C$ will form a planar graph.  In a planar
graph, if $|V|$ is the number of vertices, $|E|$ the number of edges
and $|F|$ the number of faces, then by Euler's formula
$|E|-|V|+2=|F|$. If $|F_C|$ is the number of faces of $C$ then
$|F_C|=|E_C|-|V_C|+2= |E_C|-\frac23|E_C|+2=\frac13|E_C|+2$. Thus,
$|E_C|<3|F_C|$ and hence $|V_C|=\frac23|E_C|<2|F_C|$. Or,
$|V_C|=O(|F_C|)$ and $|E_C|=O(|F_C|)$.

In two dimensions, let $S$ be a line different from $n$ given lines
(also in general position). Then $S$ will intersect (cut) some faces
of the arrangement.  Zone$(S)$ is defined as the set of faces through
which line $S$ passes. If $C\in \mbox{zone}(S)$, is a face which is cut,
then let $|C|$ be the number of edges on the boundary of face $C$ in
the (original) arrangement, then the size of zone$(S)$,
$|\mbox{zone}(S)|=\sum_{C\in \mbox{zone}(S)}|C|$. It is known that
$|\mbox{zone}(S)|=O(n)$ (\cite{R1, R2,R3,RA,ipl}).

Similarly, in three dimensions, let $S$ be a plane different from $n$
given planes (also in general position). Then $S$ will intersect (cut)
some cells of the arrangement.  Zone$(S)$ is defined as the set of
cells 
which the plane $S$ intersects. If $C\in \mbox{zone}(S)$, is a cell which is
cut, then let $|F_C|$ be the number of faces on the boundary of cell
$C$ in the (original) arrangement, then the size of zone$(S)$,
$|\mbox{zone}(S)|=\sum_{C\in \mbox{zone}(S)}|F_C|$.

Remark: Normally the size of zone is defined as the sum of number of
edges, vertices and faces of all cells in the zone, but as the number of
vertices and edges in a cell are $O(|F_i|)$, the two definitions are
equivalent up to multiplicative constants. Moreover, as each face is on
boundary of two cells, the total number of cells in a zone will also
be bounded by $O(\sum_C |F_C|)$. 

\section*{3. Zone Theorem in $3$-dimensions}

We will assume that all planes of the arrangement (together with $S$)
are in general position, as size of zone is not smaller in this case
\cite{R3,RA,R4}.

Further, let us enclose the arrangement in a ``bounding
box''\cite{R2,R3} by having six planes $x=\pm A,y=\pm A, z=\pm
A$--- basically we compute coordinates of all ${n}\choose{3}$ vertices
of arrangement (by taking every possible set of three planes) and
choosing $A$ to be larger than the (absolute value of) largest
coordinate. Thus, all cells inside the bounding box will be bounded.

Let $Q$ be any plane of the arrangement. Then as all planes are in
general position, each of them will intersect $Q$ in a line. All these
lines will lie in the plane $Q$ and form a two-dimensional (planar)
arrangement of lines (say ${\cal{L_Q}}$). 

Let us remove plane $Q$ from the arrangement ${\cal{A}}$ and let the
resulting arrangement be called ${\cal{A-Q}}$.

Let $C$ be a cell in the arrangement ${\cal{A-Q}}$. If the plane $Q$
does not cut (intersect) cell $C$, then cell $C$ and all its faces
will be present (unchanged) in arrangement ${\cal{A}}$.

If the plane $Q$ cuts (intersects) cell $C$, then the cell $C$ gets
divided into two parts--- say the part of $C$ above the plane $Q$ and
the part of $C$ below $Q$ (or if $Q$ is horizontal then left and right
of $Q$). Let us call the two parts as $C_1$ and $C_2$. Part of $C$
intersected by $Q$ will lie in plane $Q$ (definition of intersection)
and hence will be a face (say $f_Q$) in the two dimensional
arrangement ${\cal{L_Q}}$. 

If face $f$ (of $C$ in ${\cal{A-Q}}$ is not intersected by $Q$, then
face $f$ will be present (unchanged) in either $C_1$ or $C_2$.

If face $f$ is intersected by $Q$, then $f$ will get split into two
parts one above $Q$ and the other below $Q$ (or one on left and the
other on right). Let the part in $C_1$ be called $f_1$ and part in
$C_2$ be called $f_2$. Let the boundary (part common to both) be
called $e_Q$. As $e_Q$ is (also) in plane $Q$, $e_Q$ will be an edge
in two dimensional arrangement ${\cal{L_Q}}$. Edge $e_Q$ is in face
$f_Q$.

To prove the zone-theorem we need following intermediate result
\begin{theorem}
Assume ${\cal{A}}$ is an arrangement of $n$ planes, $Q$ is a plane in
${\cal{A}}$, and $S$ is a plane not in ${\cal{A}}$. Let $C$ be a cell
in zone$(S)$. Let $f$ be a face of cell $C$ not lying in plane $Q$.
Then total number of such pairs $(f,C)$ (of face $f$ and cell $C$) is
at most the sum of
\begin{enumerate}
\item size of $\mbox{zone}(S)$ in arrangement ${\cal{A-Q}}$ and
\item size of $\mbox{zone}(S)$ in two
dimensional arrangement ${\cal{L_Q}}$
\end{enumerate}
\end{theorem}

Remark: The first size is count of faces (along with their
multiplicities) and second of edges (along with their multiplicities).

\proof Assume that cell $C$ is in zone$(S)$ (of arrangement
${\cal{A-Q}}$) and $f$ is a face of $C$, not lying in (part of) plane
$Q$. As $C$ is in zone$(S)$, plane $S$ passes through cell $C$. 

Assume that cell $C$ is in zone$(S)$ (of arrangement
${\cal{A-Q}}$) and $f$ is a face of $C$, not lying in (part of) plane
$Q$. As $C$ is in zone$(S)$, plane $S$ passes through cell $C$. 

If $Q$ does not cut cell $C$, then cell $C$ (along with all its faces)
will be unchanged in arrangement ${\cal{A}}$ (and as $S$ passes
through $C$), cell $C$ will also be in zone$(S)$ in arrangement
${\cal{A}}$. In this case $f'=f$ and $C'=C$. Or the same pair is
present in both ${\cal{A}}$ and ${\cal{A-Q}}$.

Let us assume that $Q$ cuts $C$. Then cell $C$ gets divided into two
parts (say) $C_1$ and $C_2$. If the plane $Q$ does not cut face $f$,
then face $f$ will remain intact in one part (say) $C_i$ (for $i=1$ or
$2$). If part $C_i$ contains $f$, then there is one-to-one
correspondence between the pair $(f,C)$ and the pair $(f,C_i)$ (i.e.,
pair $(f,C)$ corresponds to pair $(f,C_i)$ and conversely). Note that
right hand side will be larger if $C_i$ is not in the zone (see
below).

Since $S$ passes through $C$, it will pass through either $C_1$ or
$C_2$ or both. If $S$ passes through only one part (say) $C_i$ (for
$i=1$ or $2$), then only $C_i$ will be in the zone$(S)$ in arrangement
${\cal{A}}$. In this case, for the pair $(f,C)$ we will have the
corresponding pair $(f,C_i)$ and conversely (or in case $f$ is
intersected by $Q$, then the pair $(f_i,C_i)$ where $f_i$ is the part
of $f$ in $C_i$).

We are left with the case when face $f$ is also cut by $Q$ and both
$C_1$ and $C_2$ are in the zone$(S)$.

If $S$ passes through both $C_1$ and $C_2$, then both $C_1$ and $C_2$
will be in the zone$(S)$ in arrangement ${\cal{A}}$. As $S$ passes
through both $C_1$ and $C_2$, it will also intersect the common
boundary of $C_1$ and $C_2$. But as $Q$ passes through the common
boundary of $C_1$ and $C_2$, the common part will be a face (say
$f_Q$) in the two dimensional arrangement ${\cal{L_Q}}$. And as $S$
intersects $f_Q$, face $f_Q$ will be in the two dimensional
zone$(S\bigcap Q)$.

As face $f$ is also cut by $Q$, intersection of $f$ and $Q$ will be a
line segment (say) $e_Q$. As $e$ lies in plane $Q$, $e_Q$ will be an
edge in the two dimensional arrangement ${\cal{L_Q}}$. Thus, for the
two entries $(f_1,C_1)$ and $(f_2,C_2)$ on right hand side we have two
entries: the pair $(f,C)$ in three dimensional arrangement
${\cal{A-Q}}$, and also have the pair $(e_Q,f_Q)$ in the two
dimensional arrangement ${\cal{L_Q}}$. Thus, for the two entries
$(f_1,C_1)$ and $(f_2,C_2)$ on the left hand side, we also have two
entries $(f,C)$ and $(e_Q,f_Q)$ on the right hand side.

Q.E.D.

As each face of a cell in ${\cal{A}}$ is in exactly one plane of the
arrangement, it does not lie in remaining $n-1$ planes. Thus, if we
take any pair $(f,C)$ (for face $f$ in cell $C$ lying in zone$(S)$),
it will not lie in $n-1$ planes. Or if we take each plane in turn as
plane $Q$ and add we get

$$(n-1)|\mbox{zone}(S)|\leq \sum_{Q\in {\cal{A}}}
\left(|\mbox{zone}_{\cal{A-Q}}(S)|+|\mbox{zone}_{\cal{L_Q}}(S\bigcap Q)|\right)$$

To get the bounds, let $z(n)$ be the largest possible value of
$|\mbox{zone}(S)|$ for arrangement of $n$-planes. Then, if we are 
considering this arrangement and this set $S$ (for which the value of
$|\mbox{zone}(S)|$ is the largest), then we have

$$(n-1)z(n)\leq \sum_{Q\in {\cal{A}}}
\left(|\mbox{zone}_{\cal{A-Q}}(S)|+|\mbox{zone}_{\cal{L_Q}}(S\bigcap Q)|\right)$$

As ${\cal{A-Q}}$ is an arrangement of $n-1$ planes ($Q$ is excluded),
$|\mbox{zone}_{\cal{A-Q}}(S)|\leq z(n-1)$. Further, as two dimensional
arrangement ${\cal{L_Q}}$ is in plane $Q$ and each line corresponds to
one of the other plane, the number of lines in ${\cal{L_Q}}$ is $n-1$.
By the two dimensional zone theorem (\cite{R1,
R2,R3,RA,ipl}), number of edges in zone will be linear. Hence,
for some constant $c$, $|\mbox{zone}_{\cal{L_Q}}(S\bigcap Q)|\leq c(n-1)$.
Thus, our equation becomes

$$(n-1)z(n)\leq \sum_{Q\in {\cal{A}}} \left(z(n-1)+c(n-1)\right)=
nz(n-1)+cn(n-1)$$

To solve this, we put $f(n)=z(n)/n$ or $z(n)=nf(n)$, the equation
becomes
$$f(n)\leq f(n-1)+c$$
Or $f(n)=cn$, or $z(n)=nf(n)=cn^2$. Hence we get the
Zone theorem:
\begin{theorem}
Assume that we are given an arrangement ${\cal{A}}$ of $n$ planes in
three dimension. Let $S$ be a plane different from the planes of the
arrangement. Then size of zone$(S)$, $$|\mbox{zone}(S)|=O(n^2)$$
\end{theorem}

\section*{4.$k$-Nearest Neighbours and Arrangements}

Assume $S=\left\{(x_i,y_i\right\}_{i=1}^n$ is a set of $n$ points in
the plane. Assume $p=(h,k)$ and $q=(r,s)$ are two points of $S$.

A query (or test) point $(x,y)$ will be closer to $p=(h,k)$ then to
$q=(r,s)$ iff
\begin{eqnarray*}
(x-h)^2+(y-k)^2 &<& (x-r)^2+(y-s)^2 \mbox{ or }\\
{x^2}-2hx+h^2 +{y^2}-2ky+k^2 &<& 
{x^2}-2rx+r^2 +{y^2}-2sy+s^2 \mbox{ or}\\
2rx+2sy-r^2-s^2 &<& 2hx+2ky-h^2-k^2
\end{eqnarray*}
Define $f(u,v)=2ux+2vy-u^2-v^2$

Then above condition becomes, $f(h,k)>f(r,s)$.

As equation $z=f(u,v)$ is an equation of plane, the above condition is
equivalent to saying that\\
plane $z=hx+2ky-h^2-k^2$ is above the plane $z=
2rx+2sy-r^2-s^2$.

Hence, if we draw an arrangement of $n$-planes with plane
$z=2x_ix+2y_iy-x_i^2-y_i^2$ for the $i$th point, then the nearest
neighbour of query point $(x,y)$ will be the topmost plane above
$(x,y)$ (the one visible from $(x,y,\infty)$). The second nearest
neighbout will be the second top most plane and so on. Hence, to find
$k$th closest point, we need to only consider planes which have
$(k-1)$ planes above them.

\subsection*{Acknowledgements}

I wish to thank the students who attended lectures of CS663 (2019-20
and 2024-2025) for their comments and reactions.
Thanks also to an anonymous reviewer of \cite{ipl} for pointing an
out error in the earlier version.

\bibliography{general}

\end{document}